\documentclass[aps,aps,twocolumn,superscriptaddress,showpacs,showkeys,a4paper]{revtex4}
\usepackage{dcolumn}
\usepackage{bm}
\usepackage{graphicx}
\usepackage{color}
\usepackage{subfigure}
\usepackage{hyperref}
\usepackage{latexsym}
\usepackage{amsthm}
\usepackage{amssymb}

\DeclareGraphicsExtensions{.jpg,.pdf, .mps, .png, .eps, .ps, .EPS,.gif}

\def\bc{\begin{center}} 
\def\ec{\end{center}}
\def\bea{\begin{eqnarray}}
\def\eea{\end{eqnarray}}
\newcommand{\avg}[1]{\langle{#1}\rangle}

\begin{document}

\title{Triadic closure as a basic generating mechanism of communities  in complex networks}

\author{Ginestra Bianconi}
\affiliation{School of Mathematical Sciences, Queen Mary University of London, London, UK}

\author{Richard K. Darst}
\affiliation{Department of Biomedical Engineering and Computational
  Science, Aalto University School of Science, P.O.  Box 12200,
  FI-00076, Finland}

\author{Jacopo Iacovacci}
\affiliation{School of Mathematical Sciences, Queen Mary University of
  London, London, UK}

\author{Santo Fortunato}
\affiliation{Department of Biomedical Engineering and Computational
  Science, Aalto University School of Science, P.O.  Box 12200,
  FI-00076, Finland}

\begin{abstract}
Most of the complex social, technological and biological networks have
a significant community structure. 
Therefore the community structure of complex networks has to be considered as a universal property, together with the much explored small-world and scale-free properties of these networks. Despite the large interest in characterizing the community structures of real networks, not enough attention has been devoted to the  detection of universal mechanisms able to spontaneously generate networks with communities.  Triadic closure is a
natural mechanism to make new connections, especially in social
networks. Here we show that models of
network growth based on simple triadic closure naturally lead to the emergence of community structure, together with
fat-tailed distributions of node degree, high clustering coefficients. Communities emerge from the initial stochastic heterogeneity in
the concentration of links, followed by a cycle of growth and
fragmentation. Communities are the more pronounced, the sparser the
graph, and disappear for high values of link density and randomness in
the attachment procedure. By introducing a fitness-based link
attractivity for the nodes, we find a novel phase transition, where
communities disappear for high heterogeneity of the fitness distribution, but a new
mesoscopic organization of the nodes emerges, with groups of nodes
being shared between just a few superhubs, which attract most of the links
of the system.

\end{abstract}


\maketitle
\section{Introduction}

Complex networks are characterized by a number of general properties,
that link together systems of very diverse origin, from nature,
society and technology~\cite{albert02,barrat08,newman10}. The feature that has received most attention
in the literature is the distribution of the number of neighbors of a
node (degree), which 
is highly skewed, with a tail that can be often well
approximated by a power law~\cite{albert99}. Such property explains a
number of striking characteristics of complex networks, like their high
resilience to random failures~\cite{albert00} and the very rapid
dynamics of diffusion phenomena, like epidemic
spreading~\cite{pastor01}. The generally accepted mechanism yielding
broad degree distributions is preferential
attachment~\cite{barabasi99}: in a growing network, new nodes set
links with existing nodes with a probability proportional to the
degree of the latter. This way the rate of accretion of neighbors will
be higher for nodes with more connections, and the final degrees will
be distributed according to a power law. 
Such basic mechanism, however, taken alone without considering additional growing rules, generates networks with very low values
of the clustering coefficient, a relevant feature of real
networks~\cite{watts98}. Furthermore, these networks  have no community structure~\cite{girvan02,fortunato10} either.

High clustering coefficients imply a high proportion of triads
(triangles) in the network. It has been pointed out that there is
a close relationship between a high density of triads and the
existence of community structure, especially in social networks,
where the density of triads is remarkably
high~\cite{newman03b,newman03c,toivonen06,kumpula07c,foster11}. 
Indeed, if we stick to the usual concept of communities as subgraphs with an appreciably
higher density of (internal) links than in the whole graph, one would
expect that triads are formed more frequently between nodes of the
same group, than between nodes of different
groups~\cite{granovetter73}. This concept has been actually used to
implement well known community finding
methods~\cite{palla05,radicchi04}. 
Foster et al.~\cite{foster11} have studied equilibrium graph ensembles obtained by
rewiring links of several real networks such to preserve their degree sequences
and introduce tunable values of the average clustering
coefficient and degree assortativity. They found that the modularity
of the resulting networks is the more pronounced, the larger the value
of the clustering coefficient. Correlation, however, does not imply
causation, and the work does not provide a dynamic mechanism
explaining the emergence of high clustering and community structure.

Triadic closure~\cite{rapoport53} is a strong candidate mechanism for the creation of
links in networks, especially social networks. Acquaintances
are frequently made via intermediate individuals who know both us and the
new friends. Besides, such process has the additional advantage of not
depending on the features of the nodes that get attached. With
preferential attachment, it is the node's degree that determine the
probability of linking, implying that each new node knows this information
about all other nodes, which is not realistic. Instead, triadic
closure induces an effective preferential attachment: getting linked to a neighbor $A$ of a node
corresponds to choosing $A$ with a probability increasing with the degree  $k_A$ of that node, according to a linear or sublinear
 preferential attachment. This principle is at the basis of several
generative network  models~\cite{holme02,davidsen02,vazquez03,marsili04,toivonen06,jackson07,duplication1, duplication2,duplication3,duplication4,aynaud13}, all yielding
graphs with fat-tailed degree distributions and high clustering
coefficients, as desired. Toivonen et al. have found that community
structure emerges as well~\cite{toivonen06}.

Here we propose a first systematic analysis of models based on
triadic closure, and demonstrate that this basic mechanism can indeed
endow the resulting graphs with all basic properties of real networks, including  a significant community structure.  These models can include or not an explicit preferential attachment, they can be even temporal networks, but as long as triadic closure is included, the networks are sufficiently sparse, and the growth is random, a significant  community structure spontaneously emerges in the networks. In fact the nodes of these networks are not assigned any {\em `` a priori"} hidden variable that correlates with the community structure of the networks.
 
We will first discuss a basic model including triadic closure but not an explicit preferential attachment mechanism and we will characterize the community formation
and evolution as a function of the main variables of the
linking mechanism, i.e.~the relative importance of closing a triad
versus random attachment and the average degree of the graph.  We find that communities emerge when there is a high propensity for triadic
closure and when the network is sufficiently sparse (low
average degree). We will also consider further 
models  existing in the literature and including triadic closure, and we show
that results concerning the emergence of the community structure are qualitatively the same, independently on the presence or not of the explicit preferential attachment mechanism or on the temporal dynamics of the links.
Finally, we will introduce a variant of the basic model, in
which nodes have a fitness and a propensity to attract new links
depending on their fitness. Here clusters are less pronounced
and, when the fitness distribution is sufficiently skewed, they
disappear altogether, while new peculiar aggregations of the nodes
emerge, where all nodes of each group are attached to a few superhubs. 

\section{The basic model including triadic closure}
\label{sec1}

We begin with what is possibly the simplest model of network growth
based on triadic closure. The starting point is
a small connected network of $n_0$ nodes and $m_0\geq m$ links.
The basic model contains two ingredients:
\begin{itemize}
\item{\it Growth.} At each time a new node  is added to the network
  with $m$ links.
\item{\it Proximity bias.}
The probability to attach the new node to node $i$ depends on the order in which the links are added.\\
The first link of the new node is attached to a random node $i_1$ of the network.
The probability that the new node is attached to node $i_1$ is then given by 
\begin{eqnarray}
\Pi^{[0]}(i_1)=\frac{1}{n_0+t}\,.
\end{eqnarray}
The second link  is attached to a random node of the network with probability $1-p$, while 
with probability $p$ it is attached to a node chosen randomly among the neighbors of node $i_1$.
Therefore in the first case the probability to attach to a node
$i_{2}\neq i_1$ is given by 
\begin{eqnarray}
\Pi^{[0]}(i_2)=\frac{(1-\delta_{i_1,,i_2})}{n_0+t-1}\,,
\end{eqnarray} 
where $\delta_{i_1,i_2}$ indicates the Kronecker delta,
while in the second case the probability $\Pi^{[1]}(i_{2})$ that the new node links to node $i_2$ is given by 
\begin{eqnarray}
\Pi^{[1]}({i_{2}})=\frac{  a_{i_1,i_{2}} }{k_{i_1}}\,,
\end{eqnarray}
where $a_{ij}$ is the adjacency matrix of the network and $k_{i1}$ is the degree  of node $i_1$.
\item{\it Further edges.}
For the model with $m>2$, further edges are added according to the
``second link'' rule in the previous point.  With probability $p$,
and edge is added to a random neighbor without a link of the {\it
  first} node $i_1$.  With probability $1-p$, a link is attached to
a random node in the network without a link already.  A total of $m$
edges are added, $1$ initial random edge and $m-1$ involving triadic
closure or random attachment.

\end{itemize}

\begin{figure}
\includegraphics[width=\columnwidth]{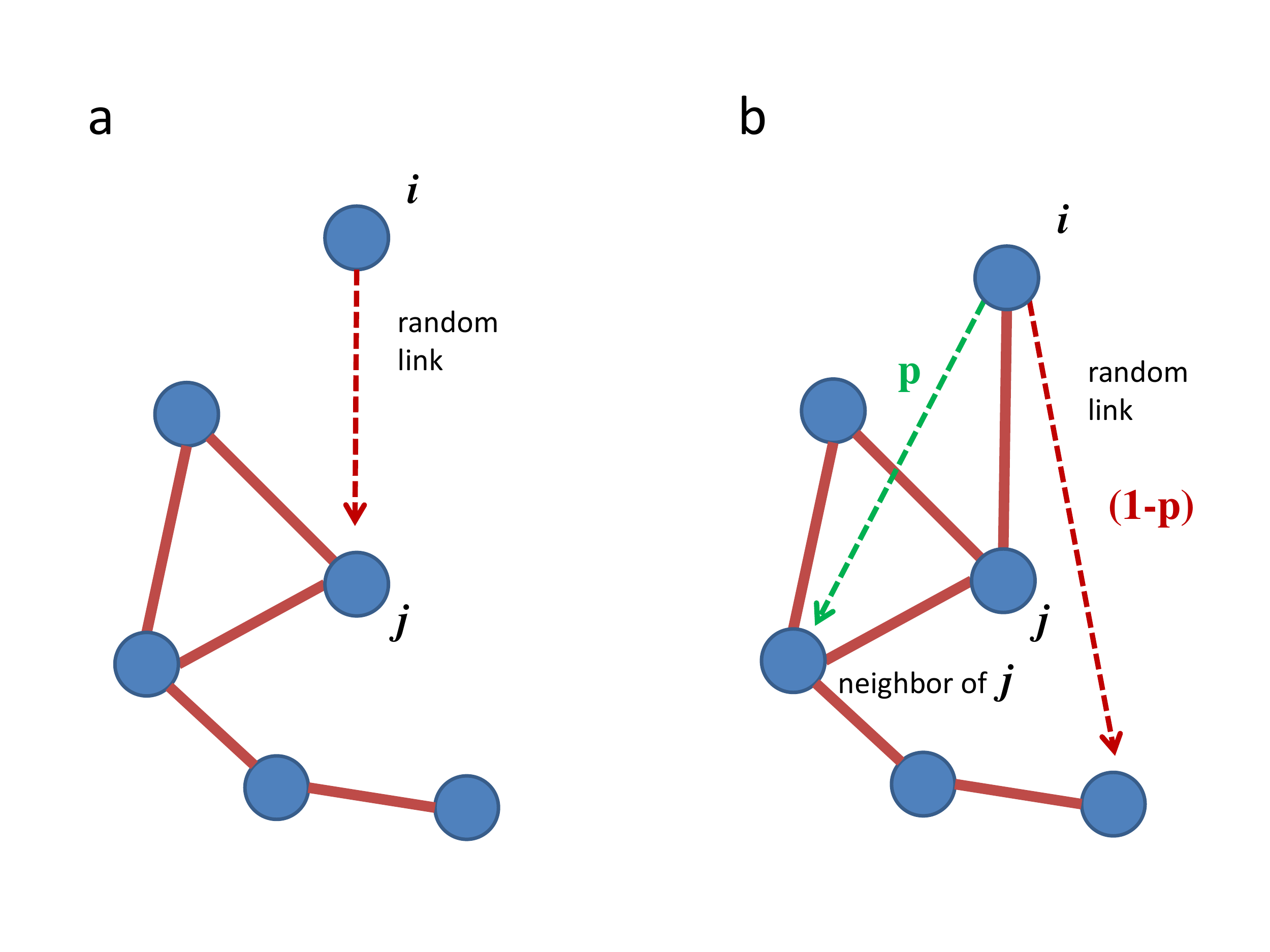} 
  \caption{(Color online) Basic model. One link associated to a new
    node $i$ is attached to a randomly chosen node $j$, the other
    links are attached to neighbors of $j$ with probability $p$,
    closing triangles, or to other randomly chosen nodes with
    probability $1-p$.}
  \label{fig1}
\end{figure}

In Fig.~\ref{fig1} the attachment mechanism of the model is
schematically illustrated.
\vspace*{0.3cm}

\begin{figure}
\includegraphics[width=\columnwidth]{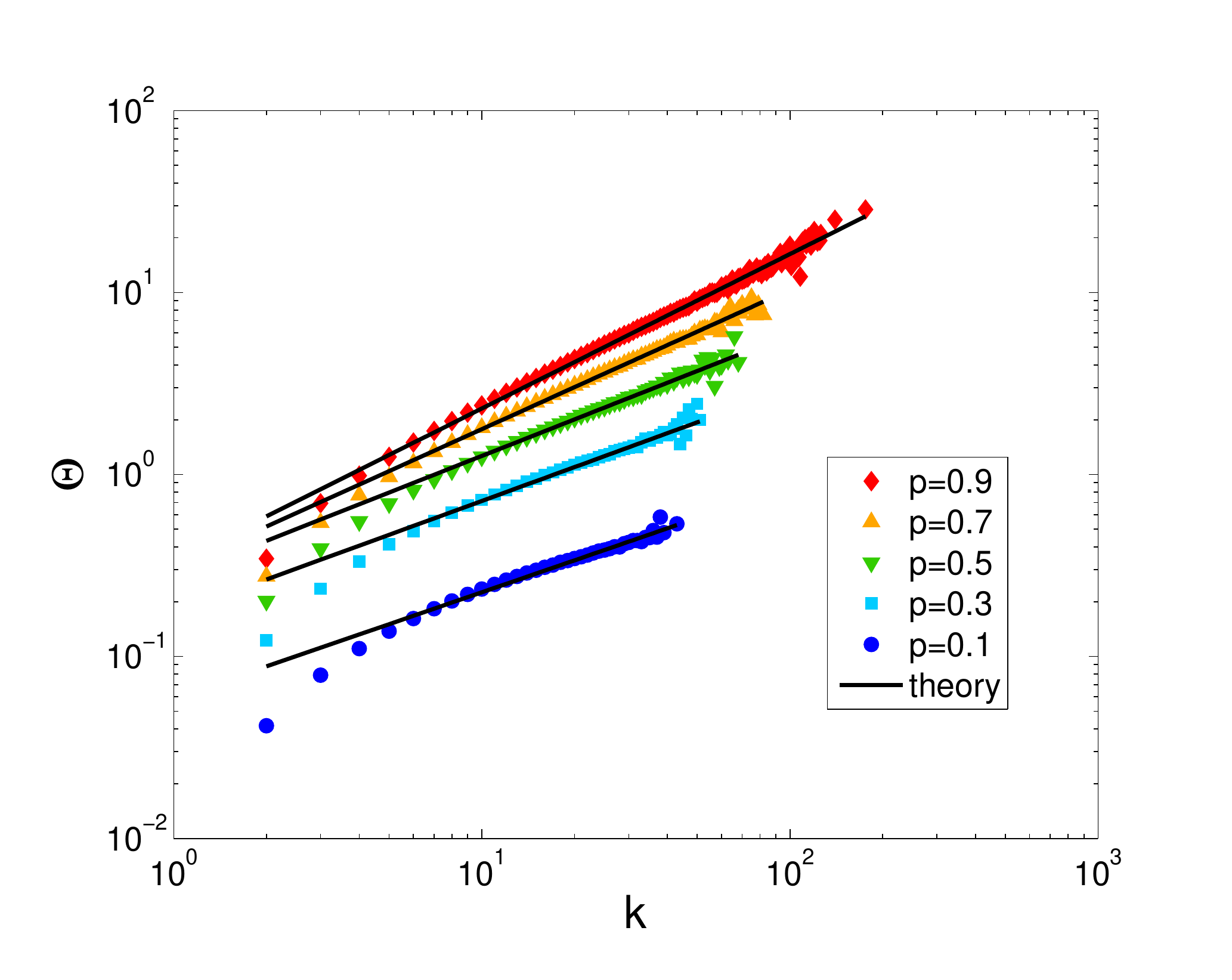} 
  \caption{(Color online) Scaling of $\Theta=\avg{\Theta_i}_{k_i=k}$,
    the average of $\Theta_i$, performed  over nodes of degree $k_i=k$,  versus  the degree $k$. This scaling allows us to define the exponents $\theta=\theta(p)$ defined by Eq. $(\ref{Theta})$. The figure is obtained by performing $100$ realizations of networks of size $n=100\,000$. }
  \label{fig:Theta}
\end{figure}

\begin{figure}
\includegraphics[width=\columnwidth]{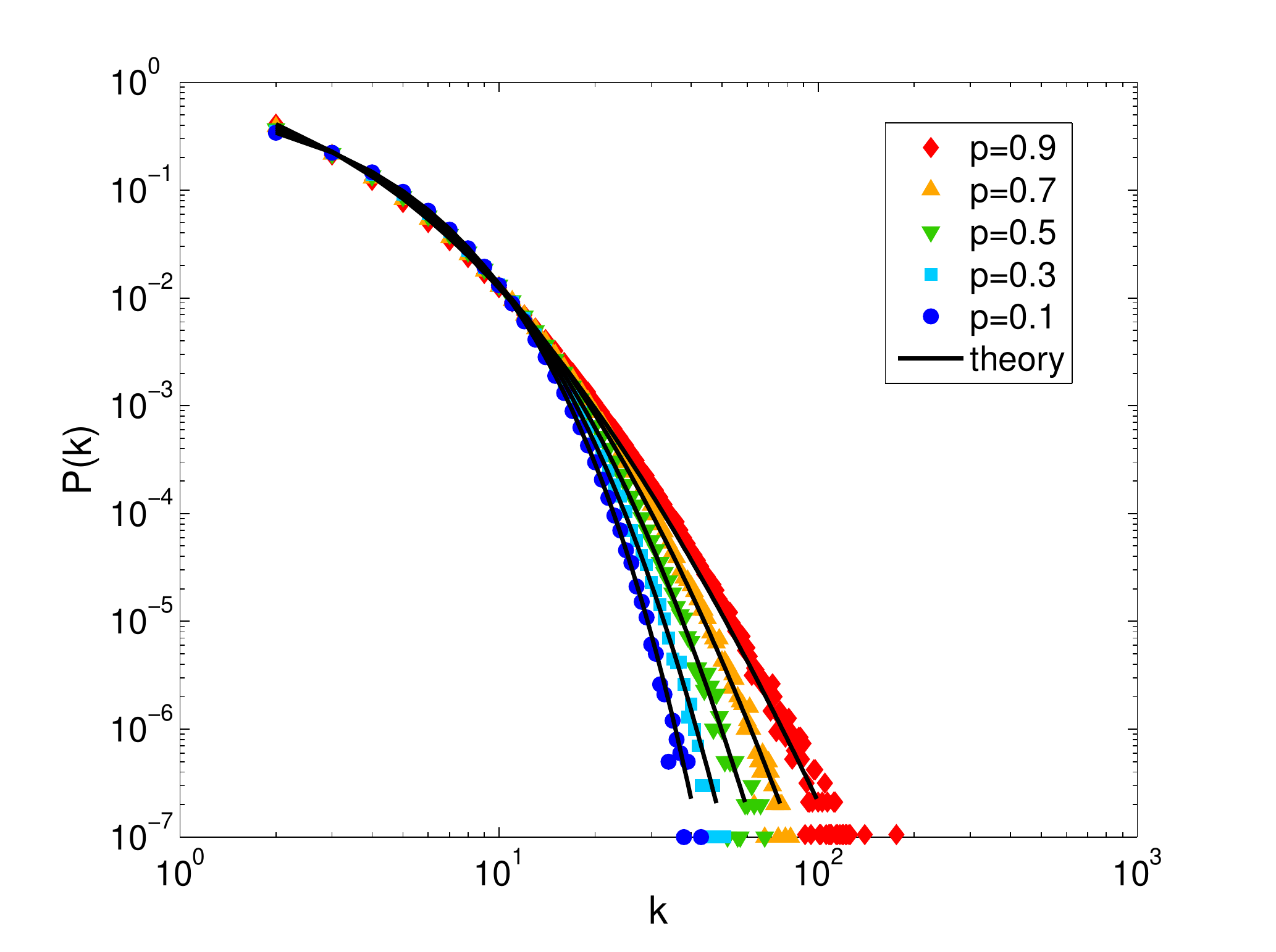} 
  \caption{(Color online) Degree distributions of the basic model, for
    different values of the parameter $p$. The continuous lines
    indicate the theoretical predictions of Eq.~$(\ref{Pk})$, the
    symbols the distributions obtained from numerical simulations of
    the model. The figure is obtained by performing $100$ realizations of networks of size $n=100\,000$.}
  \label{fig2}
\end{figure}

For simplicity we discuss here the case $m=2$.
In the basic
model the probability that a node $i$ acquires a new link at time $t$
is given by
\begin{eqnarray}
\frac{1}{t}\left[(2-p)+p\sum_j \frac{ a_{ij}}{k_j}\right].
\label{mf0}
\end{eqnarray}
In an uncorrelated network, where the probability $p_{ij}$ that a node
$i$ is connected to a node $j$ is $p_{ij}=\frac{k_ik_j}{\avg{k}n}$
($n$ being the number of nodes of the network), we might expect that the proximity bias always induces a linear preferential attachment, i.e.
\bea
\sum_j  \frac{ a_{ij}}{k_j} \propto k_i,
\eea
but for a correlated network this guess might not be correct.
Therefore, assuming, as supported by the simulation results (see Fig. ~$\ref{fig:Theta}$), that the proximity bias induces a linear or sublinear preferential attachment, i.e.
\begin{eqnarray}
\Theta_i=p\sum_j \frac{a_{ij}}{k_j}\simeq c k_i^{\theta},
\label{Theta}
\end{eqnarray}
with $\theta=\theta(p)\leq 1$ and $c=c(p)$,  we can write the master equation \cite{mendes03} for the average number $n_k(t)$ of nodes of degree $k$ at time $t$.
from the simulation results it is found that the  function $\theta(p)$ is  an increasing function of $p$ for $m=2$. Moreover the exponent $theta$ is also an increasing function of the number of edges of the new node $m$.
Assuming the scaling in Eq. $(\ref{Theta})$, the master equation for $m=2$ reads
\bea
n_{k}(t+1)&=&n_{k}(t)+\frac{2-p+c(k-1)^{\theta}}{t}n_{k-1}(t)(1-\delta_{k,2})\nonumber \\
&&-\frac{2-p+ck^{\theta}}{t}n_k(t)+\delta_{k,2}.
\eea
In the limit of large values of $t$, we assume that the degree distribution $P(k)$ can be found as $n_{k}/t\to P(k)$. So we find the solution for $P(k)$  
\begin{eqnarray}
P(k)=C\frac{1}{3-p+ck^{\theta}}\prod_{j=1}^{k-1}\left(1-\frac{1}{3-p+c j^{\theta}}\right)\,,
\end{eqnarray}
where $C$ is a normalization factor.
This expression  for $\theta<1$ can be approximated in the continuous limit by
\begin{eqnarray}
P(k)\simeq D\frac{1}{3-p+ck^{\theta}}e^{-(k-1)G(k-1,\theta,c)}\,,
\label{Pk}
\end{eqnarray}
where $D$ is the normalization constant and $G(k,\theta,c)$  is given by 
\begin{eqnarray}
G(k,\theta,c)&=&-\theta \ {_2F_1}\left(1,\frac{1}{\theta},1+\frac{1}{\theta},-\frac{ck^{\theta}}{3-p}\right)\nonumber \\
&&+\theta \ {_2F_1}\left(1,\frac{1}{\theta},1+\frac{1}{\theta},-\frac{ck^{\theta}}{2-p}\right)\nonumber \\
&&+\log\left(1-\frac{1}{3-p+ck^{\theta}}\right).
\end{eqnarray}
In this case the distribution is broad but not power law.
For $\theta=1$, instead, the distribution can be approximated in the
continuous limit by a power law, given by  
\begin{equation}
P(k)\simeq D\frac{1}{(3-p+ck)^{1/c+1}},
\label{Pkscale-free}
\end{equation}
where $D$ is a normalization constant.
Therefore we find that the network is scale free only for $\theta=1$,
i.e. only in the absence of degree correlations.
In order to confirm the result of our theory, we have extracted from the simulation results the values of the exponents $\theta=\theta(p)$ as a function of $p$.
With these values of the exponents $\theta=\theta(p)$, that turn out
to be all smaller than $1$, we have
evaluated the theoretically expected degree distribution $P(k)$ given
by Eq.~$(\ref{Pk})$ 
and we have compared it with simulations (see Fig.~\ref{fig2}), finding optimal agreement.

We remark that this model has been already studied in independent
papers by Vazquez~\cite{vazquez03} and Jackson~\cite{jackson07}, who claimed that the model yields
always power law degree distributions. Our derivation for $m=2$ shows that this is not correct, in general, and in particular it is not correct
when the growing  network exhibits degree correlations, in which case
we do not expect that the probability to reach a node of degree $k_A$
by following a link is 
proportional to $k_A$. When the network is correlated we always find
$\theta<1$, i.e. the effective link probability is {\it sublinear} in the
degree of the target node.

We note however, that the duplication model \cite{duplication1,duplication2,duplication3,duplication4}, in which every new node is attached to a random node and to each of its neighbor with probability $p$, displays at the same time degree correlations and power-law degree distribution.

We also find that the model spontaneously generates communities during
the evolution of the system. To quantify how pronounced communities
are, we use a measure called {\it embeddedness}, which estimates how
strongly nodes are attached to their own cluster. 
Embeddedness, which we shall indicate with $\xi$, is
defined as follows:
\begin{equation}
\xi=\frac{1}{n_c}\sum_c \frac{k_\mathrm{in}^c}{k_\mathrm{tot}^c}\,,
\end{equation}
where $k_\mathrm{in}^c$ and $k_\mathrm{tot}^c$ are the internal and the total degree 
of community $c$ and the sum runs over all $n_c$ communities of the network. If the
community structure is strong, most of the neighbors of each node in a
cluster will be nodes of that cluster, so $k_\mathrm{in}^c$ will be close to
$k_\mathrm{tot}^c$ and $\xi$ turns out to be close to $1$; if there is no
community structure $\xi$ is close to zero. However, one could still get values of
embeddedness which are not too small, even in random graphs, which have no modular
structure, as $k_\mathrm{in}^c$ might still be sizeable there. To eliminate
such borderline cases, we introduce a new variable, the {\it
  node-based embeddedness}, that we shall indicate with $\xi_n$. It is
based on the idea that for a node to be properly assigned to a
cluster, it must have more neighbors in that cluster than in any of
the others. This leads to the following definition 
\begin{equation}
\xi_n=\frac{1}{n}\sum_i
\frac{k_{i,\mathrm{in}}-k_{i,\mathrm{ext}}^\mathrm{max}}{k_i}\,,
\label{xin}
\end{equation}
where $k_{i,\mathrm{in}}$ is the number of neighbors of node $i$ in its
cluster, $k_{i,\mathrm{ext}}^\mathrm{max}$ is the maximum number of neighbors of
$i$ in any one other cluster and $k_i$ the total degree of $i$. The sum
runs over all $n$ nodes of the graph. For a
proper community assignment, the difference $k_{i,\mathrm{in}}-k_{i,\mathrm{ext}}^{\mathrm{max}}$
is expected to be positive, negative if the node is misclassified. In
a random graph, and for subgraphs of approximately the same size,
$\xi_n$ would be around zero. In a set of disconnected cliques (a
clique being a subgraph where all nodes are connected to each other),
which is the paradigm of perfect community structure, $\xi_n$ would
be $1$. 

\begin{figure}
\includegraphics[width=\columnwidth]{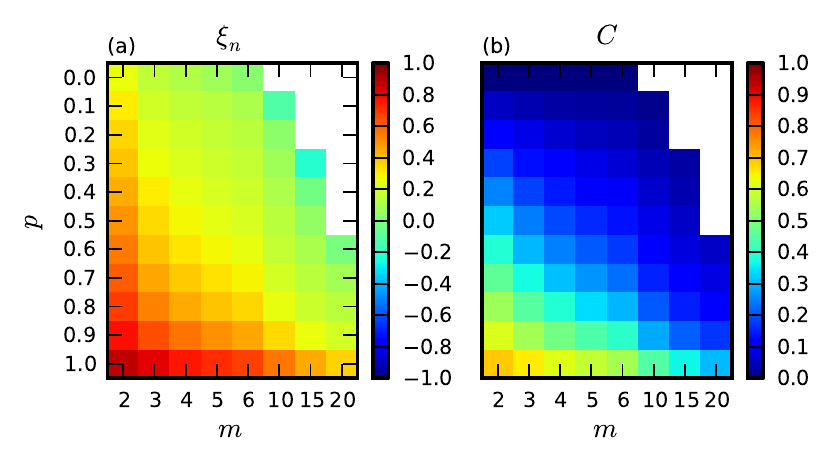} 
  \caption{(Color online) Heat map of node-based embeddedness (a) and
    average clustering coefficient (b) as a function of $p$ and $m$
    for the basic model. Community structure (higher embeddedness and clustering coefficient) is pronounced in the
    lower left region when
    $m$ is not too large (sparse graphs) and when the probability of
    triadic closure $p$ is very high. For each pair of parameter
    values we report the average over $50$ network realizations. The
    white area in the upper right corresponds to
  systems where a single community, consisting of the whole network,
  is found. Here one would get a maximum value $1$ for $\xi_n$, but it
is not meaningful, hence we discard this portion of the phase diagram, as well as in Figs. 7 and 8.}
  \label{fig3}
\end{figure}

In Fig.~\ref{fig3}a we show a heat map for $\xi_n$ as a function of the
two main variables of the model, the probability $p$ and the number of
edges per node
$m$, which is half the average degree. Communities were detected with non-hierarchical Infomap~\cite{rosvall08} in
all cases. 
Results obtained by applying the Louvain algorithm~\cite{blondel08}
(taking the most 
granular level to avoid artifacts caused by the resolution limit~\cite{fortunato07})
yield a consistent picture. All networks are grown until $n=50\,000$ nodes.
We see that large values of $\xi_n$ are associated to the bottom left
portion of the diagram, corresponding to high values of the
probability of triadic closure and to low values of degree. So, a high
density of triangles ensures the formation of clusters, provided the
network is sufficiently sparse. In
Fig.~\ref{fig3}b we present an analogous heat map for the average clustering
coefficient $C$, which is defined~\cite{watts98} as
\begin{equation}
C=\frac{1}{n}\sum_i \sum_{j,k}\frac{a_{ij}a_{jk}a_{ki}}{k_i(k_i-1)}
\end{equation}
where $a_{ij}$ is the element of the adjacency matrix of the graph and
$k_i$ is again the degree of node $i$. Fig.~\ref{fig3}b confirms that
$C$ is the largest when $p$ is high and $m$ is low, as expected.

\begin{figure}
\includegraphics[width=\columnwidth]{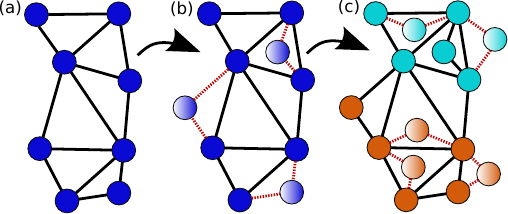} 
  \caption{(Color online) Schematic illustration of the formation and
    evolution of communities. Initial inhomogeneities in the link
    density make more likely the closure of triads in the denser
    parts, that keep growing until they become themselves inhomogeous,
  leading to a split into smaller communities (different colors).}
  \label{fig4}
\end{figure}

The mechanism of formation and evolution of communities is
schematically illustrated in Fig.~\ref{fig4}. When the first denser
clumps of the network are formed (a), out of random fluctuations in the
density of triangles newly added nodes are more likely to close triads
within the protoclusters than between them (b). 
As more nodes and links
are added, the protoclusters become larger and larger and their
internal density of links becomes inhomogenous, so there will be a
selective triadic closure within the denser parts, which yields a
separation into smaller clusters (c). This cycle of growing and
splitting plays repeatedly along the evolution of the system.
\begin{figure}
\includegraphics[width=\columnwidth]{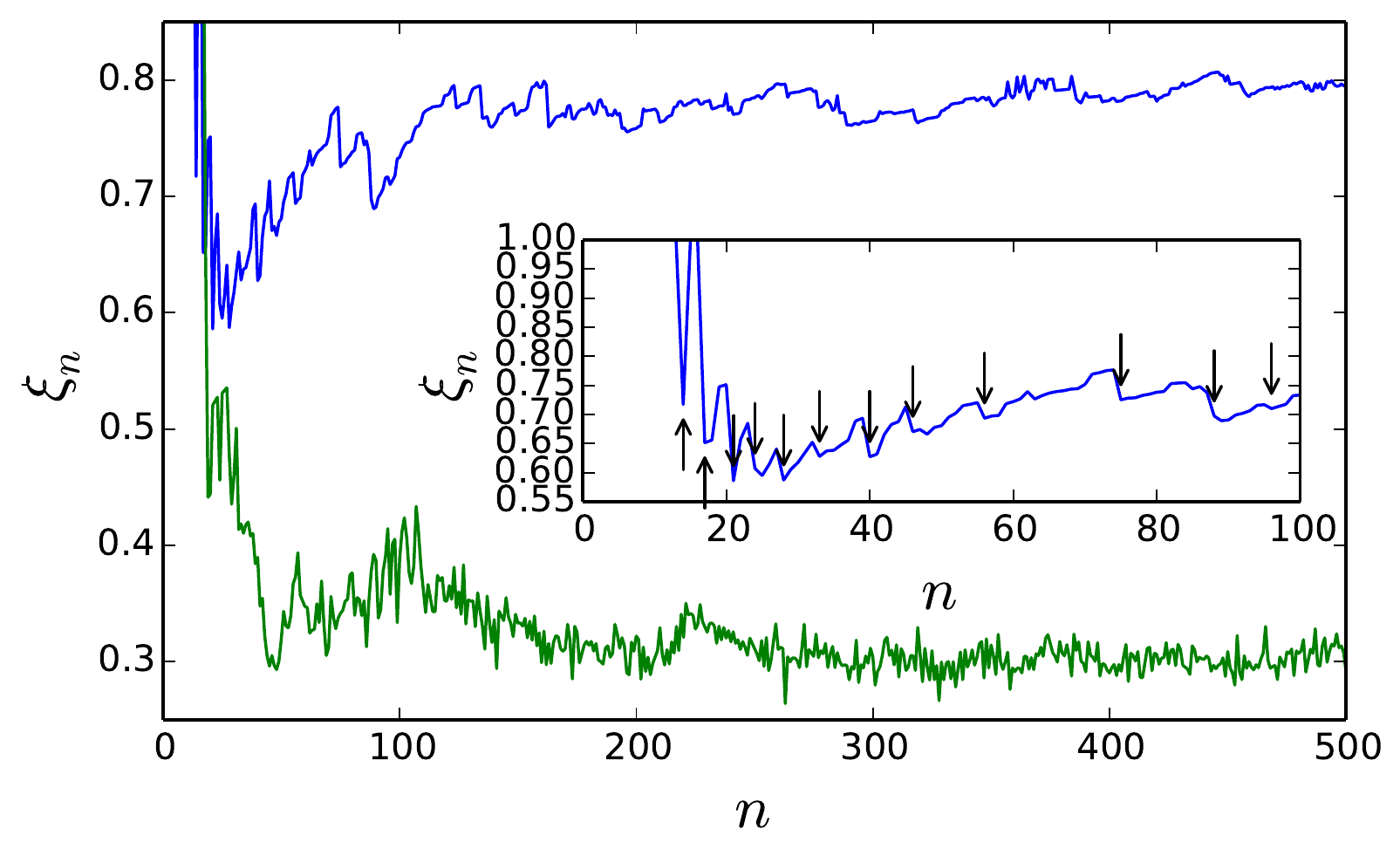} 
  \caption{(Color online) Evolution of node-based community embeddedness $\xi_n$ along
    the growth of the network. The curves refer to the extreme cases of
  absence of triadic closure (lower curve), yielding a random graph without
  communities, and of systematic triadic closure (upper curve), yielding a
  graph with pronounced community structure. For the latter case, we
  magnify in the inset the initial portion of the curve, to highlight
  the sudden drops of $\xi_n$, indicated by the arrows, which correspond to the breakout of
  clusters into smaller ones.  
}
  \label{fig5}
\end{figure}

In Fig.~\ref{fig5} we show the time evolution of the node-based embeddedness
$\xi_n$ during the growth of the system, until $500$ nodes are added to
the network, $m=2$. We consider the two extreme
situations $p=0$, corresponding to the absence of triadic closure and
$p=1$, where both links close a triangle every time and there is no
additional noise. In the first case (green line), after a transient,
$\xi_n$ sets to a low value, with small fluctuations; in the case with
pure triadic closure, instead, the equilibrium value is much higher,
indicating strong community structure,
and fluctuations are modest. In contrast with the random case, we
recognize a characteristic pattern, with $\xi_n$ increasing steadily and
then suddenly dropping. The smooth increase of $\xi_n$ signal that the
communities are growing, the rapid drop that a cluster splits into 
smaller pieces: in the inset such
breakouts are indicated by arrows. Embeddedness drops when clusters
break up because the internal degrees $k_{i,\mathrm{in}}$ of the
nodes of the fragments in
Eq.~\ref{xin} suddenly decrease, since some of the old internal
neighbors belong to a different community, while the values of
$k_{i,\mathrm{ext}}^\mathrm{max}$ are typically unaffected.

\section{Preferential attachment or temporal network models including triadic closure}

The scenario depicted in Section~\ref{sec1} is not limited to the
basic model we have investigated, but it is quite general. To show
this, we consider here two other models based on triadic closure.

The model by Holme and Kim~\cite{holme02} is a variant of the
Barab\'asi-Albert model of preferential attachment (BA model) which generate scale-free networks with clustering. The new node
joining the network sets a link with an existing node, chosen with a
probability proportional to the degree of the latter, just like in the
BA model. The other $m-1$ links coming with the new node, however, are
attached with a probability $P_t$ to a random neighbor of the node
which received the most recent preferentially-attached link,
closing a triangle, and with a
probability $1-P_t$ to another node chosen with preferential
attachment. By varying $P_t$ it is possible to tune the
level of clustering into the network, while the degree distribution is
the same as in the BA model, i.e.~a power law with exponent $-3$,
for any value of $P_t$.
\begin{figure}
\includegraphics[width=\columnwidth]{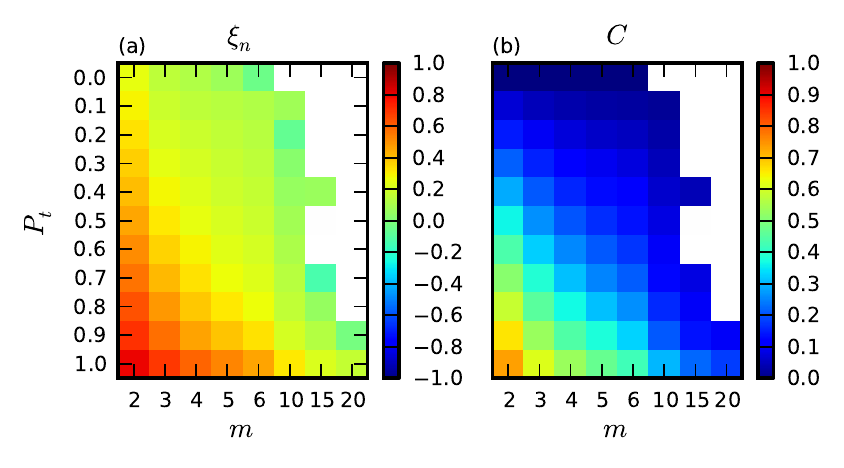} 
  \caption{(Color online) Heat map of node-based embeddedness (a) and
    average clustering coefficient (b) as a function of $P_t$ and $m$
    for the model by Holme and Kim~\cite{holme02}. For each pair of parameter
    values we report the average over $50$ network realizations. The white area in the upper right corresponds to
  systems where a single community, consisting of the whole network,
  is found, which is not interesting. The diagrams
    look qualitatively similar to that of the basic model (Fig.~\ref{fig3}),
    with highest embeddedness and clustering coefficient in the lower left region.}
  \label{fig6}
\end{figure}
In Fig.~\ref{fig6} we show the same heat map as in Fig.~\ref{fig3} for this
model, where we now report the probability $P_t$ on the
y-axis. Networks are again grown until $n=50\,000$ nodes. The
picture is very similar to what we observe for the basic model.

The model by Marsili et al.~\cite{marsili04}, at variance with most
models of network formation, is not based on a growth process.
The model is a model for temporal networks \cite{holme12}, in which the links are created and destroyed on the fast time scale while the number of nodes remains constant.  
The starting point is a random graph with $n$ nodes. Then, three
processes take place, at different rates: 
\begin{enumerate}
\item{any existing link vanishes (rate $\lambda$);}
\item{a new link is created between a pair of nodes, chosen at random
    (rate $\eta$);}
\item{a triangle is formed by joining a node with a random neighbor of
  one of his neighbors, chosen at random (rate $\xi_M$).}
\end{enumerate}

\begin{figure}
\includegraphics[width=\columnwidth]{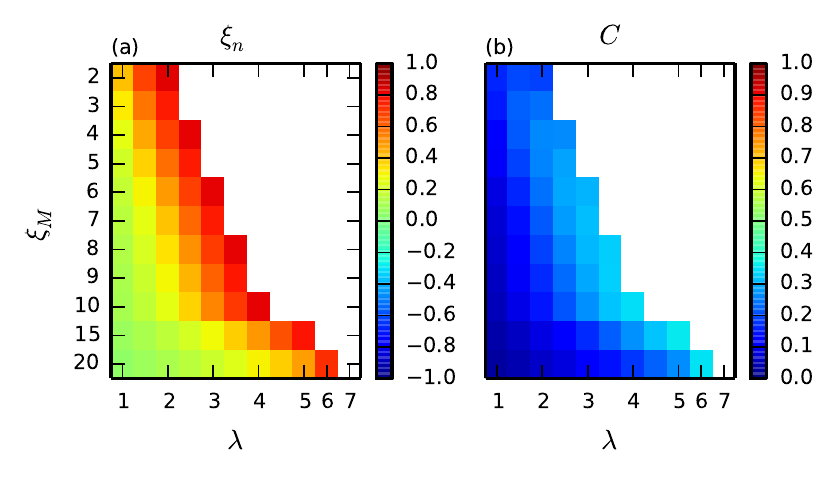} 
  \caption{(Color online) Heat map of node-based embeddedness (a) and
    average clustering coefficient (b) as a function of the rates $\lambda$ and $\xi_M$
    for the model by Marsili et al.~\cite{marsili04} ($\eta=1$). For each pair of parameter
    values we report the average over $50$ network realizations. The white area in the upper right corresponds to
  systems where a single community, consisting of the whole network,
  is found, which is not interesting. These diagrams have better communities (higher embeddedness and clustering coefficient) towards the upper right, different
  from those in Figs.~\ref{fig3} and \ref{fig6}, because of the
  different meaning and effect of the parameters. However, there is
  a strong correspondence between high clustering coefficient
  and strong community structure, as in the other models.}
  \label{fig7}
\end{figure}

In our simulations we start from a  random network of $n=50\,000$ nodes with average degree $10$. 
The three rates $\lambda$, $\eta$ and $\xi_M$ can be reduced to
two independent parameters, since what counts is their relative
size. The number of links deleted at each iteration is proportional to
$\lambda M$, where $M$ is the number of links of the network, while
the number of links created via the two other processes is
proportional to $\eta n$ and $\xi_M n$, respectively. The number of
links $M$ varies in time but in order to get a non-trivial stationary
state, one should reach an equilibrium situation where the numbers of
deleted and created links match.
A variety of scenarios are possible, depending on the choices of the
parameters. For instance, if $\xi_M$ is set equal to zero, there are
no triads, and what one gets at stationarity is a random graph with
average degree $2\eta/\lambda$. So, if $\eta\ll \lambda$, the graph is
fragmented into many small connected components. In one introduces
triadic closure, the clustering coefficient grows with $\xi_M$ if the
network is fragmented, as 
triangles concentrate in the connected components.
Moreover the model can display a veritable first order phase transition and in a region of the phase diagram displays two stable phases: one corresponding to a connected network with large average clustering coefficient and the other one corresponding to a disconnected network.  Interestingly, if there is a dense single component, the clustering coefficient decreases with
$\xi_M$. The degree distribution can follow different patterns too: it
is Poissonian in the diluted phase, where the system is fragmented, and
broad in the dense phase, where the system consists of a single
component with an appreciable density of links. In Fig.~\ref{fig7} we show the analogous heat map as in
Figs.~\ref{fig3} and \ref{fig6}, for the two parameters $\lambda$ and
$\xi_M$. The third parameter $\eta=1$. We consider only configurations where the giant component
covers more than a half of the nodes of the network. 
The diagrams are now different because of the different role
of the parameters, but the picture is consistent nevertheless.
The clustering coefficient $C$ is highest when the ratio of $\lambda$
and $\xi_M$ lies within a narrow range, yielding a sparse network with
a giant component having a high
density of triangles and a corresponding presence of
strong communities.

\section{The basic model including triadic closure and  fitness of the nodes}

In this Section we introduce a variant of the basic model, where the
link attractivity depends on some intrinsic fitness of the nodes. 
We will assume that the nodes are not all equal and assign to each node $i$ a fitness $\eta_i$ representing the ability of a node to attract new links.
We have chosen to parametrize the fitness with a parameter $\beta>0$ by setting 
\begin{eqnarray}
\eta_i=e^{-\beta \epsilon_i}\,,
\end{eqnarray}
with $\epsilon$ chosen from a distribution $g(\epsilon)$ and $\beta$ representing a tuning parameter of the model.
We take 
\begin{eqnarray}
g(\epsilon)=(1+\nu)\epsilon^{\nu}\,,
\label{ge}
\end{eqnarray}
with $\epsilon \in (0,1)$.
When $\beta=0$ all the fitness values are the same, when $\beta$ is large small differences in the $\epsilon_i$ cause large differences in fitness.
For simplicity we assume that the fitness values are quenched variables
assigned once for all to the nodes. As in the basic model without fitness, the starting point is
a small connected network of $n_0$ nodes and $m_0\geq m$ links.
The  model contains two ingredients:
\begin{itemize}
\item{\it Growth}. At time $t$ a new node is added to the network with
  $m\ge 2$ links. 
\item{\it Proximity and fitness bias}.
The probability to attach the new node to node $i_1$ depends on the order in which links are added.\\
The first link  of the new node is attached to a random node $i_1$ of the network with probability proportional to its fitness.
The probability that the new node is attached to node $i_1$ is then given by 
\begin{eqnarray}
\Pi^{[0]}({i_1})=\frac{\eta_{i_1}}{\sum_{j}\eta_j}\,.
\end{eqnarray}
For $m=2$ the second link is attached to a node of the network chosen according
to its fitness, as above, with probability $1-p$, while 
with probability $p$ it is attached to a node chosen randomly between the neighbors of the node $i_1$ with probability proportional to its fitness.
Therefore in the first case the probability to attach to a node $i_{2}\neq i_1$ is given by 
\begin{eqnarray}
\Pi^{[0]}(i_2)=\frac{\eta_{i_2}(1-\delta_{i_1,i_2})}{\sum_{j\neq i_1}\eta_j}\,,
\end{eqnarray} 
with $\delta_{i_1,i_2}$ indicating the Kronecker delta,
while in the second case the probability $\Pi^{[1]}({i_{2}})$ that the new node links to node $i_2$ is given by 
\begin{eqnarray}
\Pi^{[1]}(i_{2})=\frac{\eta_{i_2} a_{i_1,i_2} }{\sum_{j}\eta_{j}a_{i_1,j}}\,,
\end{eqnarray}
where  $a_{ij}$ indicates the matrix element $(i,j)$ of the adjacency matrix of the network.

\item{\it Further edges}.
For $m>2$, further edges are added according to the
``second link'' rule in the previous point.  With probability $p$
an edge is added to a neighbor of the {\it first}
node $i_1$, not already attached to the new node, according to the fitness rule.  With probability $1-p$, a
link is set to any node in the network, not already
attached to the new node, according to the fitness rule.  
\end{itemize}

For simplicity we shall consider here the case $m=2$.  The probability
that a node $i$ acquires a new link at time $t$ is given by
\begin{equation}
\frac{e^{-\beta\epsilon_{i}}}{t}\left[(2-p)+p\sum_j \frac{ a_{ij}}{\sum_{r}\eta_r a_{jr}}\right]\,.
\end{equation}
Similarly to the case without fitness, here we will assume, 
supported by simulations, that  
\begin{equation}
\Theta_i=p\sum_j \frac{\eta_j a_{ij}}{\sum_{r} \eta_r a_{jr}}\simeq ck_i^{\theta(\epsilon)}\,,
\end{equation}
where, for every value of $p$, $\theta=\theta(\epsilon)\leq1$ and $c=c(\epsilon)$. 

We can write the master equation for the average number $n_{k,\epsilon}(t)$ of nodes of degree $k$ and energy $\epsilon$ at time $t$, as
\begin{eqnarray}
n_{k,\epsilon}(t+1)&=&n_{k,\epsilon}(t)\nonumber \\
&&\hspace*{-15mm}+\frac{e^{-\beta\epsilon}[2-p+c(\epsilon)(k-1)^{\theta}]}{t}n_{k-1,\epsilon}(t)(1-\delta_{k,2})\nonumber \\
&&\hspace*{-15mm}-\frac{e^{-\beta\epsilon}[2-p+c(\epsilon)k^{\theta(\epsilon)}]}{t}n_{k,\epsilon}(t)+\delta_{k,2}g(\epsilon)\,.
\end{eqnarray}

\begin{figure}
\includegraphics[width=\columnwidth]{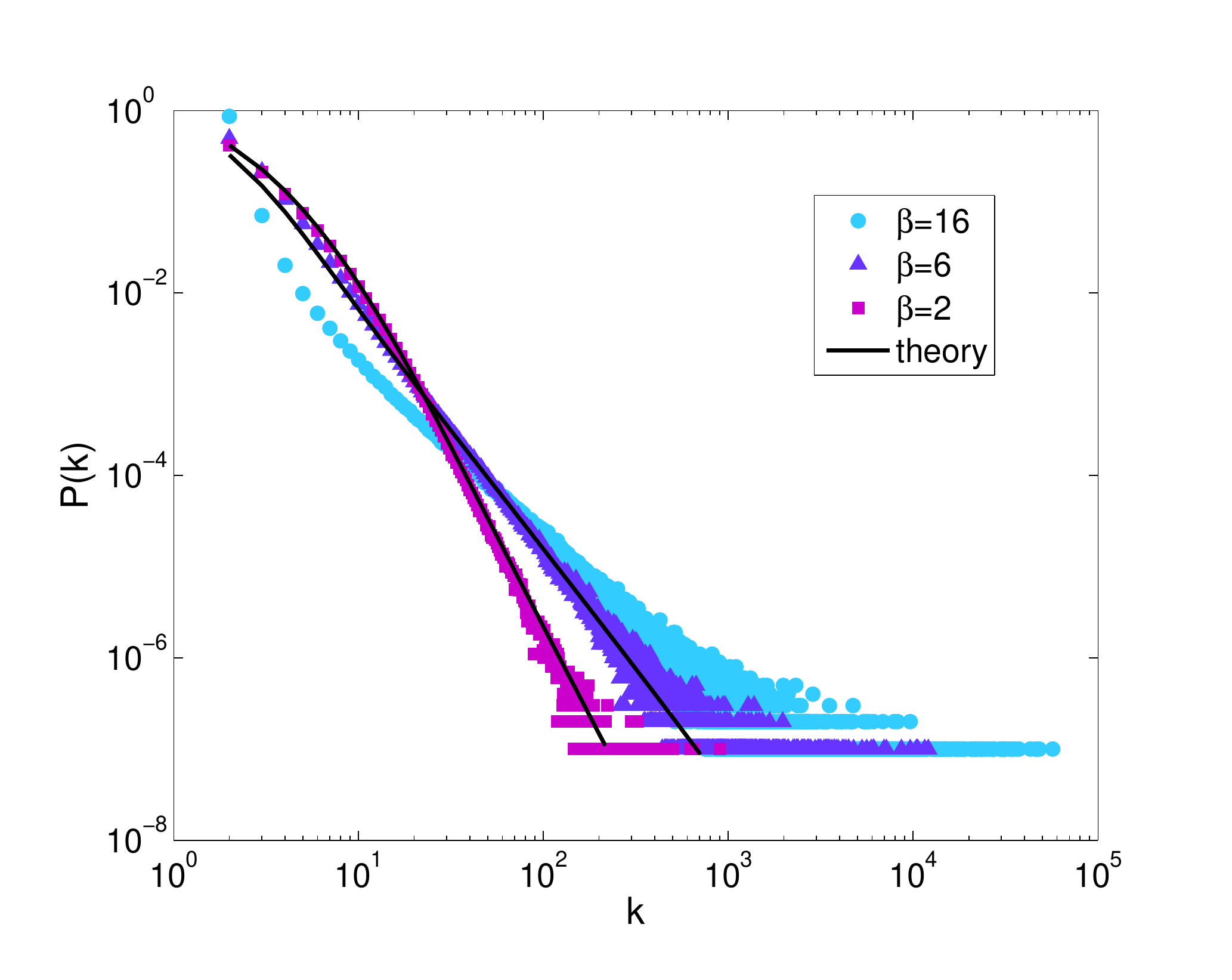} 
  \caption{(Color online) Degree distribution of the model with
    fitness, for three values of the parameter $\beta$, which
    indicates the heterogeneity of the distribution of the 
    fitness of the nodes. Symbols stand for the results
    obtained by building the network via simulations, continuous lines
  for our analytical derivations. The figure is obtained by performing $100$ realizations of networks of size $n=100\,000$ with  $\nu=6$.}
  \label{fig8}
\end{figure}

In the limit of large values of $t$ we assume that
$n_{k,\epsilon}/t\to P^{\epsilon}(k)$, and therefore we find that the solution for $P^{\epsilon}(k)$ is given by

\begin{eqnarray}
P^{\epsilon}(k)&=&C(\epsilon)\frac{1}{1+e^{-\beta\epsilon}[2-p+c(\epsilon)k^{\theta(\epsilon)}]}\nonumber \\&&
\hspace*{-7mm}\times\prod_{j=1}^{k-1}\left\{1-\frac{1}{1+e^{-\beta\epsilon}[2-p+c(\epsilon) j^{\theta(\epsilon)}]}\right\}\,,
\end{eqnarray}
where $C(\epsilon)$ is the normalization factor.
This expression  for $\theta(\epsilon)<1$ can be approximated in the continuous limit by 
\begin{eqnarray}
P^{\epsilon}(k)&\simeq &D(\epsilon)\frac{e^{-(k-1)G[k-1,\epsilon,\theta(\epsilon),c(\epsilon)]}}{1+e^{-\beta\epsilon}[2-p+c(\epsilon)k^{\theta(\epsilon)}]},
\end{eqnarray}
where $D(\epsilon)$ is the normalization constant and $G(k,\epsilon,\theta,c)$ is given by 
\begin{eqnarray}
G(k,\epsilon,\theta,c)&=&-\theta _2F_{1}\left(1,\frac{1}{\theta},1+\frac{1}{\theta},-\frac{ck^{\theta}}{2-p+e^{\beta\epsilon}}\right)\nonumber \\
&& +\theta _2F_{1}\left(1,\frac{1}{\theta},1+\frac{1}{\theta},-\frac{ck^{\theta}}{2-p}\right)\nonumber \\
&&+\log\left(1-\frac{1}{1+\frac{e^{\beta\epsilon}}{2-p+ck^{\theta}}}\right)\,.
\end{eqnarray}
When $\theta(\epsilon)=1$, instead, we can approximate
$P^{\epsilon}(k)$ with a power law, i.e.
\begin{equation}
P^{\epsilon}(k)\simeq D(\epsilon)\left[1+e^{-\beta \epsilon}\left(2-p+c(\epsilon) k\right)\right]^{-\frac{e^{-\beta\epsilon}}{c(\epsilon)}-1}.
\end{equation}
Therefore, the  degree distribution $P(k)$ of the entire network is a convolution of  the degree distributions $P^{\epsilon}(k)$ conditioned on the value of $\epsilon$, i.e.
\begin{eqnarray}
P(k)=\int d\epsilon P^{\epsilon}(k)\,.
\label{Pkfit}
\end{eqnarray}
As a result of this expression, we found that the degree
distribution can be a power law also if the network exhibits degree correlations and $\theta(\epsilon)<1 $ for every value of $\epsilon$. Moreover we observe that for  large values of the parameter $\beta$ the distribution becomes broader and broader until a condensation transition occurs at $\beta=\beta_c$ with the value of $\beta_c$ depending on both the parameters $\nu$ and $p$ of the model. For $\beta>\beta_c$  successive nodes with maximum fitness (minimum value of  $\epsilon$) become ``superhubs'', attracting a finite fraction of all the links, similarly to what happens in Ref.~\cite{bianconi01}.
In Fig.~\ref{fig8} we see the degree distribution of model, obtained
via numerical simulations, for different values of $\beta$. The
continuous lines, illustrating the theoretical behavior, are well
aligned with the numerical results, as long as $\beta<\beta_c$.

\begin{figure}
\includegraphics[width=\columnwidth]{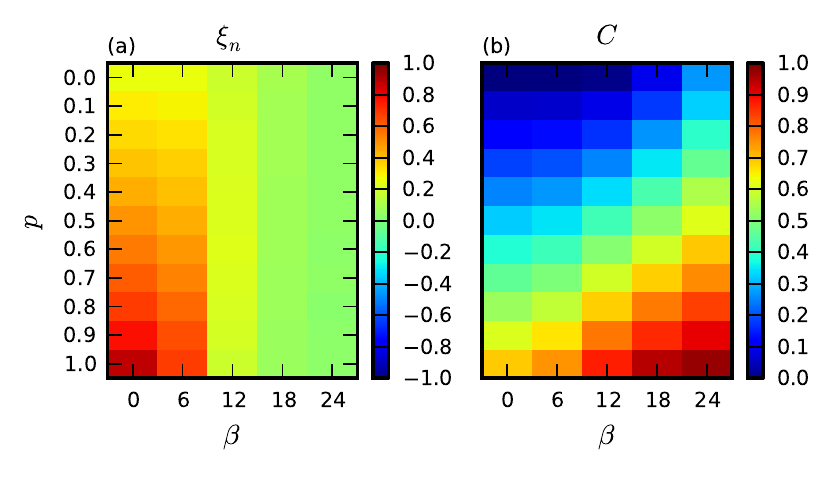} 
  \caption{(Color online) Heat map of node-based embeddedness (a) and
    average clustering coefficient (b) as a function of the
    probability of triadic closure $p$ and the heterogeneity parameter
    $\beta$ of the fitness distribution of the nodes, for the model
    with fitness. The number of new edges per node is $m=2$. For each pair of parameter
    values we report the average over $50$ network realizations.
    When $\beta=0$ we
    recover the basic model, without fitness.   We see
    the highest values of embeddedness in the lower left, while highest
    values of the clustering coefficient are in the lower right.
    When $\beta$ increases,
    we see a drastic change of structure in contrast to the previous pattern: communities disappear,
    whereas the clustering coefficient gets higher.}
  \label{fig9}
\end{figure}

In Fig.~\ref{fig9} we show the heat map of $\xi_n$ and $C$ for the
model, as a function of the parameters $p$ and $\beta$. The number of edges per node
is $m=2$, and the networks consist of $50\,000$ nodes. Everywhere in
this work, we set the parameter $\nu=6$.
For $\beta =0$ all nodes have identical fitness and the
model reduces itself to the basic model. So we recover the previous
results, with the emergence of communities for sufficiently large
values of the probability of triadic closure $p$, following a 
large density of triangles in the system. The situation changes
dramatically when $\beta$ starts to increase, as we witness a
progressive weakening of community structure, while the clustering
coefficient keeps growing, which appears counterintuitive. In the
analogous diagrams for $m=5$, we see that this pattern holds, though
with a weaker overall community structure and lower values of the
clustering coefficient.
\begin{figure}
\includegraphics[width=\columnwidth]{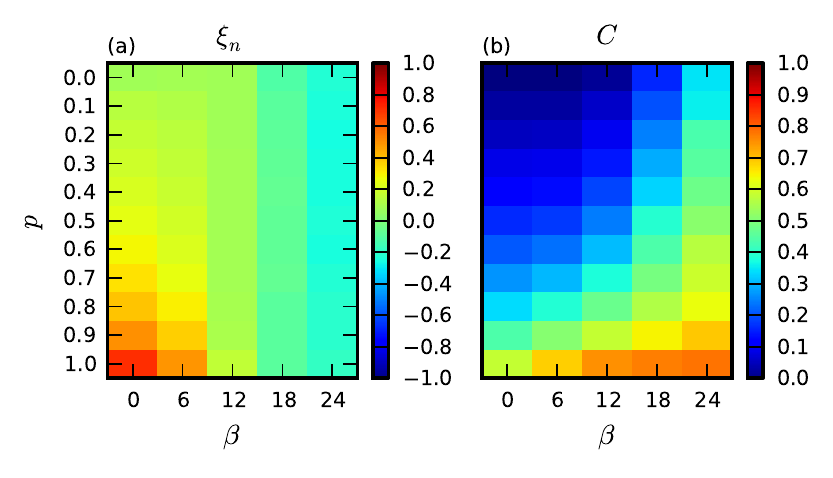} 
  \caption{(Color online) Same as Fig.~\ref{fig9}, but for $m=5$. The
  picture is consistent with the case $m=2$, but communities are less pronounced.}
  \label{fig10}
\end{figure}

\begin{figure*}
\includegraphics[width=\textwidth]{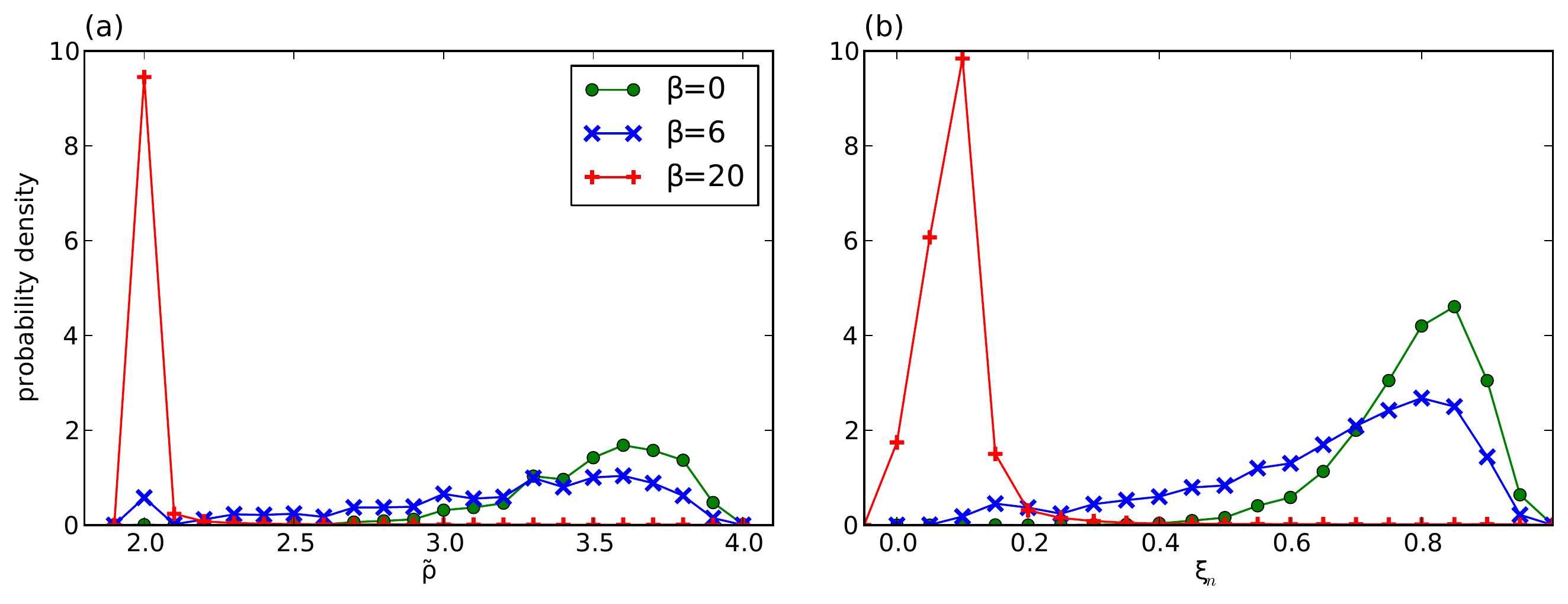} 
  \caption{(Color online) Probability distributions of 
the scaled link density $\tilde{\rho}$ (left) and node-based embeddedness
    $\xi_n$ (right) of the communities of the
fitness model, for $m=2$ and $\beta= 0, 6, 20$. For each $\beta$-value we
derived $100$ network realizations, each with $100\,000$ nodes.  We see that
at $\beta=0$, the detected communities satisfy the expectations of good communities,
while at $\beta=20$ they do not.}
  \label{fig11}
\end{figure*}

When $\beta$ is sufficiently large, communities disappear, despite the
high density of triangles. To check what happens, we compute the
probability distribution of 
the scaled link density $\tilde{\rho}$ and the node-based embeddedness $\xi_n$ of the communities of the
networks obtained from $100$
runs of the model, for three different values of $\beta$: $0$, $6$ and
$20$. All networks are grown until $100\,000$ nodes. The scaled link density
$\tilde{\rho}$ of a cluster is defined~\cite{lancichinetti10b} as
\begin{equation}
\tilde{\rho}=\frac{2l_c}{n_c-1}\,,
\end{equation}
where $l_c$ and $n_c$ are the number of internal links and of nodes of
cluster $c$. If the cluster is tree-like, $\tilde{\rho}\approx 2$, if
it is clique-like it $\tilde{\rho}\approx n_c$, so it grows linearly
with the size of the cluster. The distributions of $\xi_n$ and $\tilde{\rho}$
are shown in Fig.~\ref{fig11}. 
They are peaked, but the peaks undergo a
rapid shift when $\beta$ goes from $0$ to $20$. The situation
resembles what one usually observes in first-order phase transitions. The embeddedness ends
up peaking at low values, quite distant from the maximum $1$, while
the scaled link density eventually peaks sharply at $2$, indicating
that the subgraphs are effectively tree-like.

\begin{figure*}
\includegraphics[width=\textwidth]{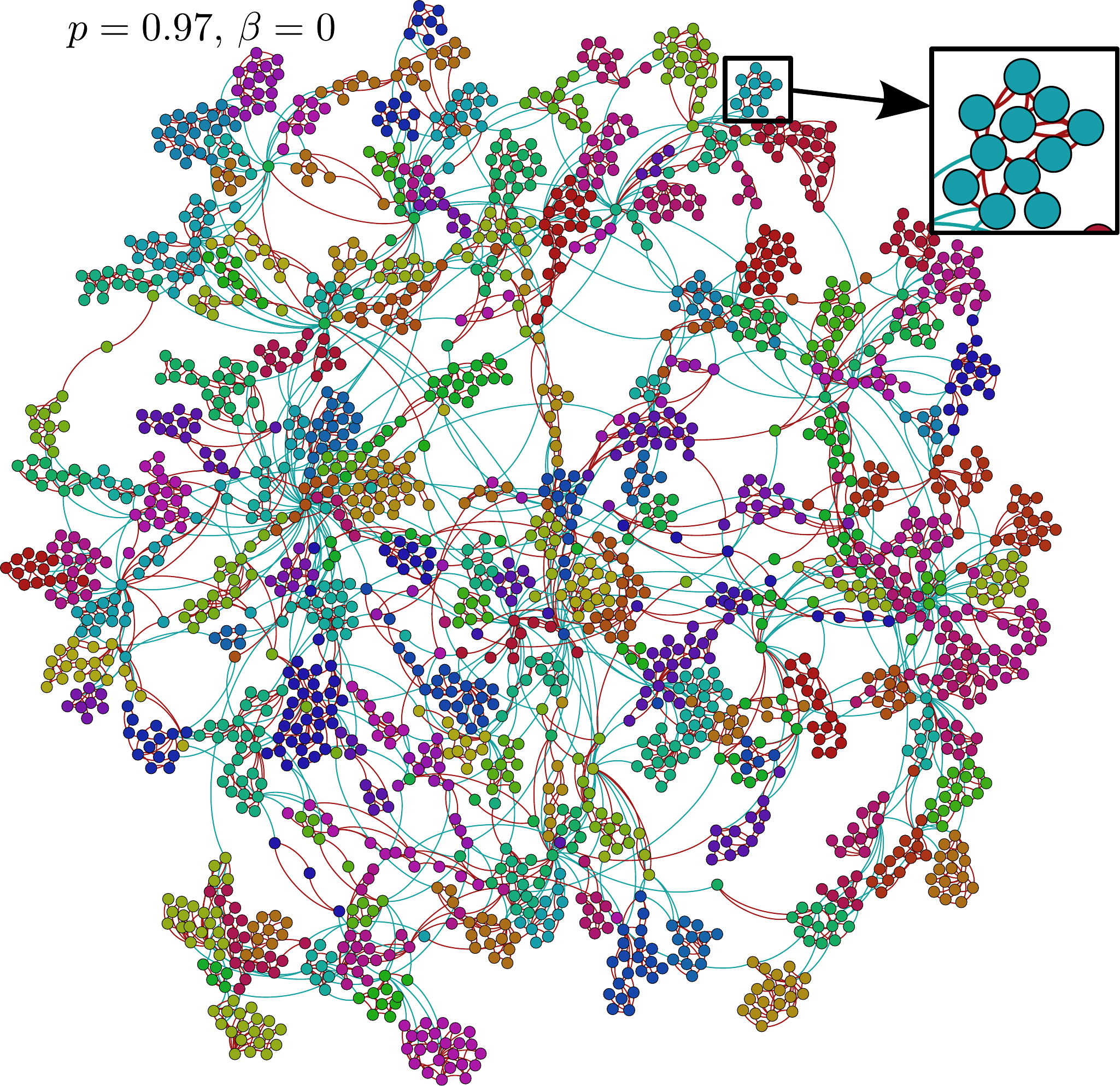} 
  \caption{(Color online) Picture of a network with $2\,000$ nodes generated by the
    fitness model, for $p=0.97$, $m=2$ and $\beta=0$.  Since $\beta=0$
    fitness does not play a role and we recover the results of the
    basic model. 
    Colors indicate communities as detected by the non-hierarchical Infomap
    algorithm~\cite{rosvall08}.}
  \label{fig12}
\end{figure*}

\begin{figure*}
\includegraphics[width=\textwidth]{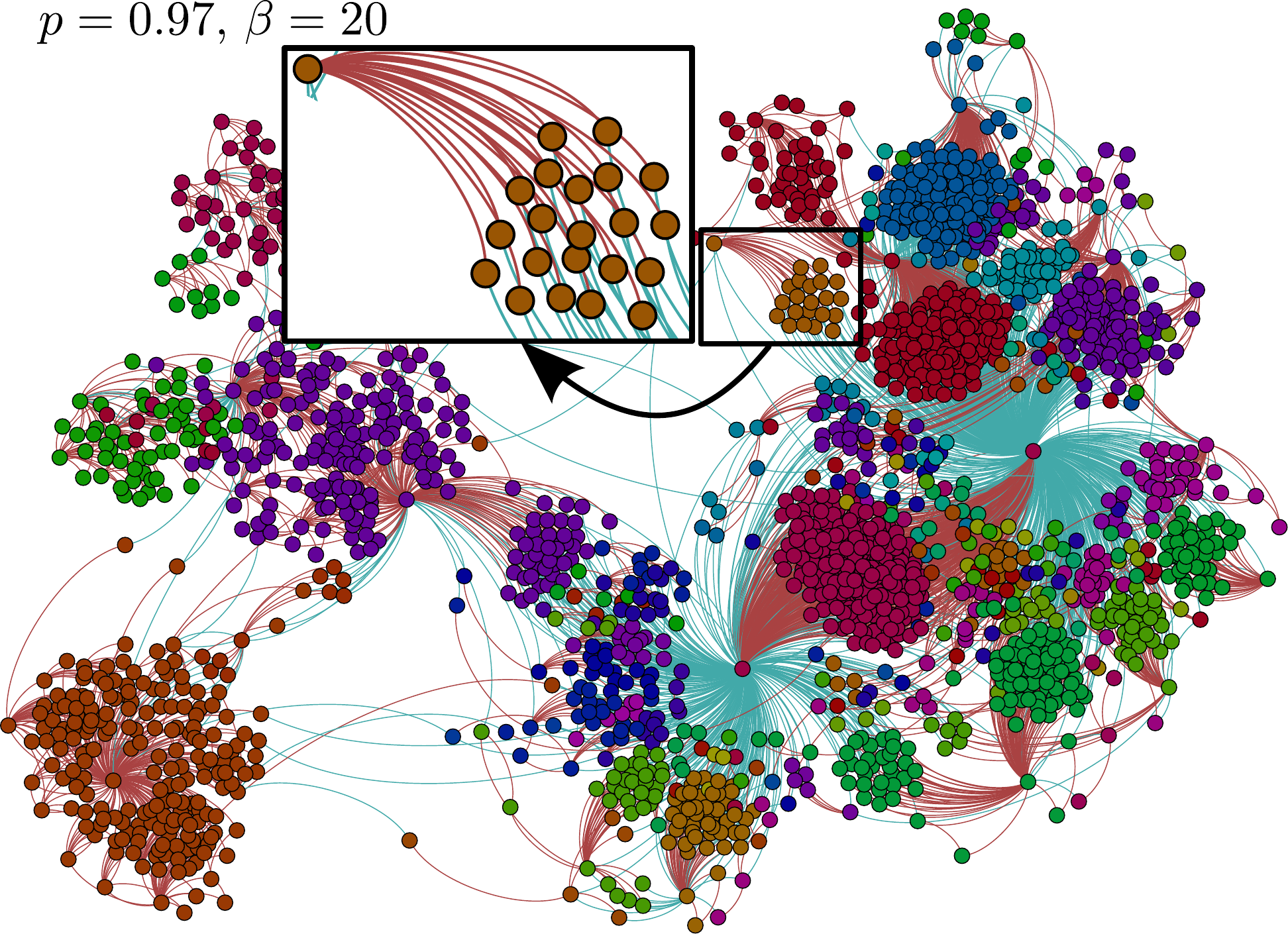} 
\caption{(Color online) Picture of a network with $2\,000$ nodes
  generated by the fitness model, for $p=0.97$, $m=2$ and $\beta=20$.
  The growing process is the same as in Fig.~\ref{fig12}, but the
  addition of fitness changes the structural organization of the network.  As
  seen in the inset, node aggregations form around hub nodes with
  high fitness.  Looking at the inset we see that such aggregations
  do not satisfy the typical requirements for communities:
  they are internally tree-like, and there are more external edges
  (blue or light gray) than internal (red or dark gray) touching its nodes. In particular,
  internal edges only go from regular nodes to superhubs.}
  \label{fig13}
\end{figure*}

What kind of objects are we looking at? To answer this question, in
Figs.~\ref{fig12} and \ref{fig13} we display two pictures of networks
obtained by the fitness model, for $\beta=0$ and $\beta=20$,
respectively. The number of nodes is $2\,000$, and the number of edges per node
$m=2$. The probability of triadic closure is $p=0.97$, as we
want a very favorable scenario for the emergence of
structure. 
The subgraphs found by our community detection method (non-hierarchical Infomap,
but the Louvain method yields a similar picture) are identified by the
different colors. The insets show an enlarged picture of the
subgraphs, which clarify the apparent puzzle delivered by the previous
diagrams. For the basic model $\beta=0$ (Fig.~\ref{fig12}), the
subgraphs are indeed communities, as they are cohesive objects which
are only loosely connected to the rest of the graph. The situation
remains similar for low values of $\beta$. However, for sufficiently
high $\beta$ (Fig.~\ref{fig13}), a phenomenon of link condensation
takes place, with a few superhubs attracting most of the links of the
network~\cite{bianconi01}. Most of the other nodes are organized in
groups which are ``shared''
between pairs (for $m=2$, more generally $m$-ples) of superhubs (see
figure).
The community embeddedness is low because there are always many links
flowing out of the subgraphs, towards superhubs. 
Besides, since the superhubs are all linked to each other, this
generates high clustering coefficient for the subgraphs, as observed
in Figs.~\ref{fig9} and \ref{fig10}. In fact, the
clustering coefficient for
the non-hubs attains the maximum possible value of $1$, as their neighbors
are nodes which are all linked to each other.

\section{Conclusions}

Triadic closure is a fundamental mechanism of link formation,
especially in social networks. We have shown that such mechanism alone
is capable to generate systems with all the characteristic properties
of complex networks, from fat-tailed degree distributions to high
clustering coefficients and strong community structure. In particular,
we have seen that communities emerge naturally via triadic closure, 
which tend to generate cohesive subgraphs around portions of the
system that happen to have higher density of links, due to stochastic
fluctuations. When clusters become sufficiently large, their internal
structure exhibits in turn link density inhomogeneities, leading to a
progressive differentiation and eventual separation into smaller
clusters (separation in the sense that the density of links between
the parts is appreciably lower than within them).
This occurs both in the basic version of network growth model based on
triadic closure, and in more complex variants. The strength of
community structure is the higher, the sparser the network and the
higher the probability of triadic closure.

We have also 
introduced a new variant, in that link attractivity depends on some
intrinsic appeal of the nodes, or fitness. Here we have seen that,
when the distribution of fitness is not too heterogeneous, community
structure still emerges, though it is weaker than in the absence of
fitness. By increasing the heterogeneity of the fitness distribution,
instead, we observe a major change in the structural organization of
the network: communities disappear and are replaced by special
subgraphs, whose nodes are connected only to superhubs of the network,
i.e.~nodes attracting most of the links. Such structural phase
transition is associated to very high values of the
clustering coefficient.

\begin{acknowledgments}
R. K. D. and S. F. gratefully acknowledge MULTIPLEX, grant number
317532 of the European Commission and the computational resources
provided by Aalto University Science-IT project.
\end{acknowledgments}

\end{document}